\documentclass[10pt,conference]{IEEEtran}
\IEEEoverridecommandlockouts

\usepackage{cite}
\usepackage{amsmath,amssymb,amsfonts}
\usepackage{algorithmic}
\usepackage{graphicx}
\usepackage{textcomp}
\usepackage{xcolor}

\usepackage{enumitem}
\usepackage{amsmath}
\usepackage[hypertexnames=false,pdfborder={0 0 0},hidelinks]{hyperref}
\usepackage{breakurl}
\usepackage{lscape}
\usepackage{threeparttable}
\usepackage{multirow}
\usepackage{array}
\usepackage{tablefootnote}
\usepackage{multicol}
\usepackage{ascmac}
\usepackage{fancybox}
\usepackage{footnote}
\usepackage{here}

\def\BibTeX{{\rm B\kern-.05em{\sc i\kern-.025em b}\kern-.08em
    T\kern-.1667em\lower.7ex\hbox{E}\kern-.125emX}}
\begin{document}

\title{Generating High-Level Test Cases from Requirements using LLM: An Industry Study}

\author{\IEEEauthorblockN{Satoshi Masuda}
\IEEEauthorblockA{\textit{Tokyo City University}\\
Kanagawa, Japan\\
\ smasuda@tcu.ac.jp}
\and
\IEEEauthorblockN{Satoshi Kouzawa, Kyousuke Sezai, Hidetoshi Suhara, \\ Yasuaki Hiruta, Kunihiro Kudou}
\IEEEauthorblockA{\textit{VeriServe Corporation}} 
Tokyo, Japan \\
\{satoshi.kouzawa, kyousuke.sezai, hidetoshi.suhara, \\ yasuaki.hiruta, kunihiro.kudou\}@veriserve.co.jp}

\maketitle

\begin{abstract}
Currently, generating high-level test cases described in natural language from requirement documents is performed manually. In the industry, including companies specializing in software testing, there is a significant demand for the automatic generation of high-level test cases from requirement documents using Large Language Models (LLMs). Efforts to utilize LLMs for requirement analysis are underway. In some cases, retrieval-augmented generation (RAG) is employed for generating high-level test cases using LLMs. However, in practical applications, it is necessary to create a RAG tailored to the knowledge system of each specific application, which is labor-intensive. Moreover, when applying high-level test case generation as a prompt, there is no established method for instructing the generation of high-level test cases at a level applicable to other specifications without using RAG. It is required to establish a method for the automatic generation of high-level test cases that can be generalized across a wider range of requirement documents. In this paper, we propose a method for generating high-level (GHL) test cases from requirement documents using only prompts, without creating RAGs. In the proposed method, first, the requirement document is input into the LLM to generate test design techniques corresponding to the requirement document. Then, high-level test cases are generated for each of the generated test design techniques. Furthermore, we verify an evaluation method based on semantic similarity of the generated high-level test cases. In the experiments, we confirmed the method using datasets from Bluetooth and Mozilla, where requirement documents and high-level test cases are available, achieving macro-recall measurement of 0.81 and 0.37, respectively. We believe that the method is feasible for practical application in generating high-level test cases without using RAG.
\end{abstract}


\section{Introduction}
\subsection{Background}
In the field of software engineering, extensive research has been conducted on test automation \cite{Shamshiri2015, 5477060}. Notably, advancements have been made in the automation of unit tests and the automation of test data input operations into web browsers, which have been practically implemented \cite{7965450}. These fall under the category of test execution automation. On the other hand, when creating high-level (HL) test cases such as system tests and user acceptance tests, it is common to derive test cases from requirement documents \cite{9440190}. 
A high-level test case is described in natural language, for instance, as 'Verify that several bookmarks are deleted at one time'. Currently, the creation of high-level test cases is often performed manually and is frequently documented in natural language. This task depends on the engineer's abilities, which may lead to omissions in test cases. Consequently, advanced research on the automation of test case creation documented in natural language has been conducted \cite{9103626}. Recently, research utilizing large language models (LLMs) for test case generation has also emerged \cite{10440574}. Beyond test case generation, many proposals in text generation using LLMs employ retrieval-augmented generation (RAG) techniques \cite{10628480}. This technique involves pre-storing data from the target knowledge domain and adding this data to user inputs when using the language model, thereby achieving more accurate outputs.

Companies specializing in software testing are also engaged in automating the generation of test cases documented in natural language. This research targets the testing of a wide range of software, including accounting systems and logistics systems used in corporate back offices, software embedded in home appliances, and software installed in automobiles. To accommodate such diverse software, it is necessary to prepare retrieval-augmented generation tailored to each business when applying retrieval-augmented techniques. However, preparing them individually requires business-specific adaptations, and establishing a common retrieval-augmented technique makes it extremely challenging to determine what should be common. Therefore, we aim to automate the creation of high-level test cases documented in natural language without using retrieval-augmented techniques. This is the motivation for our research.

\begin{figure*}[h!]
\centering
 \includegraphics[width=180mm,bb=0 0 951 309]{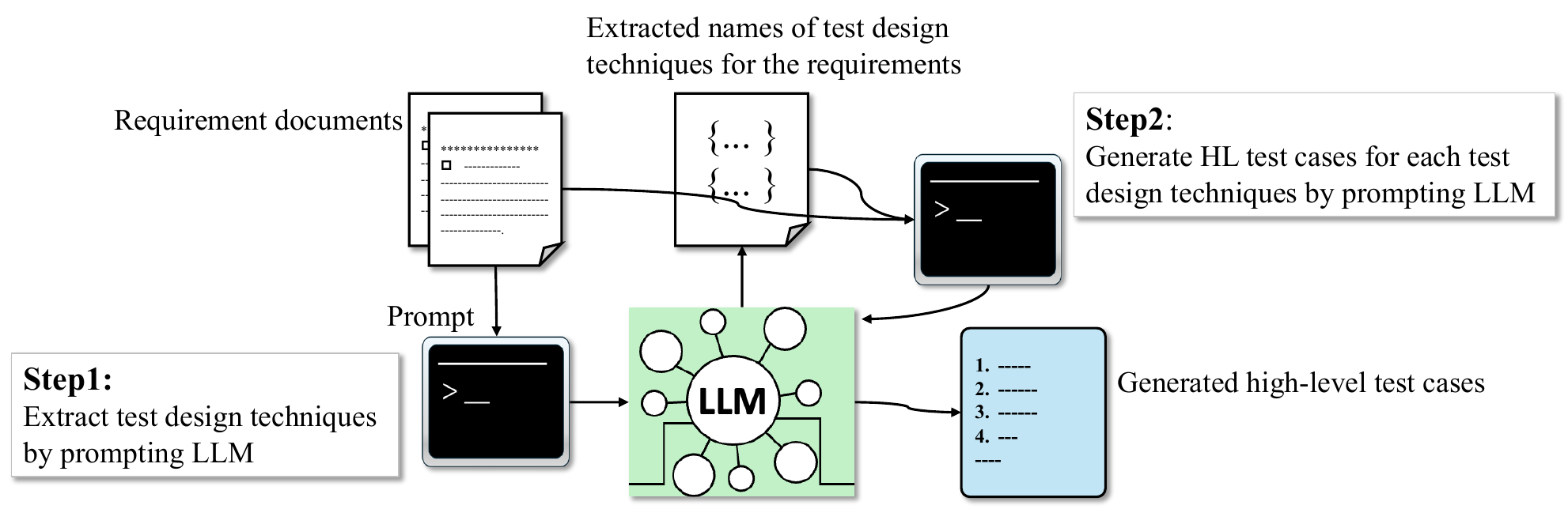}
 \caption{Overview of proposed method: Generating High-level test cases using LLM (GHL)}
\label{figure:overview}
\end{figure*}

\subsection{Challenges}
In practical, creating RAG itself is labor-intensive and its accuracy is unstable \cite{2402.19473}. RAG utilizes structured business-specific knowledge to enhance search effectiveness. When applying test case creation instructions as prompts to establish a method for automatically generating test cases that can be generalized to more requirement documents, there is no established method for creating test case instructions at a level applicable to other specifications without using RAG. Therefore, this study investigates a method for generating test cases from requirement documents using only prompts, without creating RAG.

When outputs are generated automatically, a method to verify the correctness of these outputs is necessary. This research focuses on test cases documented in natural language. The truth test cases are also documented in natural language. Specifically, the test cases targeted in this research are referred to as high-level test cases, which describe the test overview. Therefore, it is necessary to compare sentences documented in natural language. Although there are many methods for comparing sentences documented in natural language, examples of their use in the technical comparison of software test cases are few. Thus, confirming whether the evaluation method proposed in this research is effective for test case comparison is crucial. Determining the correctness of automatically generated test cases documented in natural language is important.

This paper clarifies the following research questions:

\begin{itemize}
\item RQ1: {\bf What prompts are effective for generating high-level test cases from requirement documents using LLMs with only prompts?}
\item RQ2: {\bf How can the truthfulness of the generated high-level test case descriptions be evaluated?}
\item RQ3: {\bf What are the currently resolved points and future challenges for the practical application of automatic high-level test case generation from requirements?}
\end{itemize}

By elucidating these aspects, this study contributes to the establishment of test case generation techniques from requirements using LLMs within the industrial sector.

This paper envisions the automation of high-level test case generation and the subsequent automation of test execution as a follow-up process. However, the scope of this study is limited to investigating the accuracy and efficiency of high-level test case generation. The specific methods for automation are not addressed in this paper.

\subsection{Proposed Method and Contribution}
We propose a {\bf g}enerating {\bf h}igh-{\bf l}evel test cases method ({\bf GHL}) from requirement documents using {\bf L}LM with only prompts, without creating RAG.
In the proposed method, first, test design techniques corresponding to the requirement documents are generated by inputting the requirement documents into the LLM. Next, high-level test cases are generated for each generated test design technique.

Fig. \ref{figure:overview} shows overview of the proposed method. We report experimental results of high-level test case generation using industrial data. We report the evaluation method for the high-level test case generation method. This contributes to the software industry.

\subsection{Structure of This Paper}
In Chapter II, we provide a summary of related researches. This chapter discusses existing technologies related to LLMs and software testing. Chapter III provides a detailed explanation of the proposed method. In Chapter IV, we describe the experiments conducted to validate the proposed method. This chapter summarizes the experimental methods, the datasets used, and the experimental environment. Chapter V presents the experimental results and their evaluation, followed by a discussion in Chapter VI. Finally, we conclude and outline future works in chapter VII.

\section{Related works}
\subsection{LLM and Prompt Engineering}

Currently, LLMs \cite{NEURIPS2020_1457c0d6} are beginning to be widely used in the field of software engineering, including requirements engineering. Particularly, the use of LLMs is advancing in document analysis and code analysis. In requirements engineering, there are related works utilizing LLMs for test case generation \cite{10628513,10628464,10429996}. 
In these studies, examples of using RAG to improve accuracy are common. RAG is a mechanism that improves the accuracy of outputs from LLMs by adding pre-stored business knowledge to prompts. While RAG contributes to accuracy improvement, the challenge lies in how to accumulate business knowledge. We believe that it is extremely difficult to prepare RAG for all test case generation in practical terms. Many studies have already been conducted on what inputs should be given to obtain desired results from LLMs. As a result of this prompt engineering, it has been clarified that providing inputs step-by-step yields more accurate desired results \cite{10.5555/3600270.3601883}. In this paper, we aim to generate more accurate test cases by providing test case generation tasks to LLMs step-by-step.

\subsection{Retrieval Augmented Generation (RAG)}

RAG is a technique that combines the generative capabilities of LLMs with external knowledge bases or document retrieval systems for tasks such as question answering and document generation \cite{10.5555/3495724.3496517}. 
This method aims to provide more accurate and up-to-date information by utilizing external data rather than relying solely on the knowledge retained within the model. 
The mechanism of RAG involves first receiving the user's question or request as input and searching for relevant documents from external databases or search engines. 
Then, based on the search results (relevant documents), the generative model generates answers or text. The LLM generates appropriate responses to the user by referring to the information obtained from the search. 
The advantage of RAG is that it allows the use of the latest information and incorporates knowledge that emerged after the model's training. 
Additionally, by expanding the knowledge base, it is possible to update knowledge by simply changing or updating external databases without retraining the model. 
Furthermore, there is no need to encompass large amounts of knowledge, allowing for a reduction in model size. 
Examples of RAG applications include question-answering (QA) systems, document retrieval and summarization, customer support chatbots, and legal document search and analysis \cite{10.5555/3495724.3496517}. 

However, RAG has several challenges and issues. 
If the search results from external databases are inaccurate or incomplete, the generated results may be adversely affected \cite{10852457}. 
If the search engine fails to properly understand the user's intent, there is a risk of incorporating less relevant information. 
For instance, insufficient keyword matching may return information that differs from the question's intent. 
If the database is insufficiently updated, the accuracy of time-dependent questions (e.g., latest news or technology trends) may decrease. 
For instance, application in frequently updated fields such as medicine or law is challenging \cite{10822677, 10661194}. 
Due to these challenges, we are investigating methods for generating test cases without using RAG.

\subsection{Software Testing}
Software testing is a crucial process in software development. To create test cases, it is necessary to thoroughly read requirement documents and design test cases based on them. Various test design techniques are used for creating these test cases. Representative techniques include equivalence partitioning, boundary value analysis, and decision table testing \cite{Myers.2012}. For instance, decision table testing begins by decomposing requirements documented in natural language into conditions and actions. Based on these conditions, a decision table is created. The decision table is completed by describing how actions change according to the truth values of the conditions. Next, specific values are input for the truth values of the conditions. By detailing how actions result from these specific conditions, test cases using the decision table testing technique are created. 
In this workflow, the creation of test cases by humans typically requires several minutes to several tens of minutes per test case. For instance, it may take approximately 120 minutes to create 30 test cases. At the Fraunhofer Institute in Germany, the time required for humans to create test cases is estimated to be between 5 to 10 minutes per test case. This estimation is used to evaluate the effectiveness of test case generation using LLMs \cite{fraunhofer202502}.

These test design techniques have been systematized through previous research and are also compiled as international standards. The international standard ISO/IEC/IEEE 29119 Part 4 \cite{9591574} details test design techniques. The ISO/IEC/IEEE 29119 Part 4 classifies various test techniques as follows:
\begin{itemize}
\item Specification-based (black-box) test techniques
\begin{itemize}
\item Equivalence Partitioning
\item Boundary Value Analysis
\item State Transition Testing
\item Decision Table Testing
\item Combinatorial Testing, etc.
\end{itemize}
\item Structure-based (white-box) test techniques
\begin{itemize}
\item Control Flow Testing
\item Data Flow Testing
\item Branch Testing
\item Statement Coverage, etc.
\end{itemize}
\item  Experience-based test techniques
\begin{itemize}
\item Exploratory Testing
\item Error Guessing, etc.
\end{itemize}
\end{itemize}

These classifications provide guidance for selecting appropriate methods based on the characteristics and objectives of test techniques. Based on these test design techniques, engineers create test cases by leveraging their knowledge. Specifically, techniques such as boundary value analysis and combinatorial testing are used to generate test cases from requirement documents. This task is dependent on the engineer's skills and is subject to individual capabilities. Therefore, research on automating test case creation using natural language has been conducted, but no practically usable methods have been established at present.

Test strategy in the international standard ISO/IEC/IEEE 29119 \cite{9698145, 9591577, 9340095} provides high-level guidance on how to conduct an organization's testing activities. The test strategy functions as a framework for determining what tests to conduct, which test techniques to use, how to allocate resources, and how to prioritize tests to achieve the organization's quality goals. It is desirable for the test strategy to be documented and often used in conjunction with test policies and test plans. It is particularly effective for unifying testing activities in large-scale projects or across an organization. When creating test cases from requirement documents, software engineers must follow the test strategy established by the organization. Therefore, the content of the test strategy is extremely important information when creating test cases.

\section{Proposed Method}

Fig. \ref{figure:overview} shows steps of the proposed method.
In the proposed method: GHL, we first input the requirement document into a LLM to generate test design techniques tailored to the requirement document. Subsequently, for each generated test design technique, we produce high-level test cases.

In the Step 1 of generating test design techniques, both the requirement document and the test strategy are used as inputs to create test design techniques based on the requirement document. Multiple test design techniques can be generated in this process, including, for instance, equivalence partitioning and boundary value analysis. These test design techniques are appropriately generated according to the requirement document and the test strategy. The extracted test design techniques are generated by the LLM based on the requirement documents and test strategy. These are not merely lists of test design techniques but are appropriately selected test design techniques according to the importance of the requirements and tests. 

In the Step 2, for each extracted test design technique, specific high-level test cases are generated from the requirement documents. Each high-level test case is a test case description created based on test design techniques such as equivalence partitioning and boundary value analysis. Therefore, the generated test cases are created from various test design perspectives, increasing comprehensiveness. Fig.\ref{figure:overview} illustrates the process of test case generation using LLMs. Below is a detailed explanation of each part. The input part is the requirement document described in natural language. This is a specification document written in text, serving as the basis for creating test cases. 

Additionally, the test strategy document is also used as input. Together with these requirement documents and test strategy, questions are given to the LLM to extract suitable test design techniques. The extracted test design techniques include documents related to multiple test design techniques. Representative methods include equivalence partitioning, boundary value analysis, and state transition testing. Furthermore, the LLM  is a model that receives natural language specifications and test design methods as input and generates test cases. The generation process is conducted by combining the knowledge of test design methods with the content of the specifications. As a result, high-level test cases \cite{1224448} are generated. These are test cases generated by the LLM. "High-level" indicates that specific test data or procedures may not be included, and rather, test scenarios or conditions are expected to be described. 

Moreover, Fig.\ref{figure:overview} provides an overview of the workflow for test case generation using LLMs. By combining specification documents described in natural language with test design methods, this approach efficiently generates test cases. Particularly, by providing test design methods in advance, it demonstrates an effort to enable the LLM to generate appropriate test cases.

\section{Experiment}
\subsection{Experiment Overview}
We conducted an evaluation of test case generation accuracy using an existing dataset (Bluetooth, Mozilla) that includes requirements and test cases. The evaluation involved method of generating test cases using a {\bf zero-shot} prompt (only questions for test case creation) and the proposed method: {\bf GHL} and its variation GHL-function ({\bf GHL-F}). GHL-F is in addition to GHL, test case generation includes combinations of {\bf f}unctions.

{\bf zero-shot}: The zero-shot approach generates test cases with a single prompt.
\begin{itembox}[l]{zero-shot}
You are software testing expert. 
The attached file1 shows requirements of a system.
The attached file2 shows test strategy of the system. 
Can you extract as much as possible test cases from the requirements, 
including normal cases and abnormal cases in the test strategy?
\end{itembox}

{\bf GHL}: Test cases are created for each test design technique by providing the requirements, including various scenarios such as test control and hands-free headsets.

\begin{itembox}[l]{GHL -  Step 1:Extract test design techniques}
The attached file shows requirements of the system.
Can you extract as much as possible candidate test design techniques in ISO/IEC/IEEE 29119-4:2021 for the requirements?
If you can't extract the test design techniques, extract as much as possible popular test design techniques in the ISO/IEC/IEEE 29119-4:2021 in general.
\end{itembox}

For each extracted testing technique, test cases are generated. The {\verb|<|\bf test design\verb|>|} as below is replaced for name of each test design techniques. 
\begin{itembox}[l]{GHL -  Step 2: for each test design techniques}
The attached file shows requirements of the system. Extract test cases as much as possible, according with the
{\verb|<|\bf test design\verb|>|}
technique including normal cases and abnormal cases in the test strategy.
\end{itembox}

\begin{table*}[htbp]
 \centering
\begin{threeparttable}[h]
 \caption{Information of the datasets: Bluetooth and Mozilla}
 \label{table:datasets}
  \begin{tabular}{l|lrrp{5em}|rp{5em}}
   \hline
 &  \multicolumn{4}{c|}{Requirements}  & \multicolumn{2}{c}{HL Test cases (Truth)}  \\ \cline{2-7}
Datasets & Function & Pages & Words & URL (*)& \begin{tabular}{r} Number of high\\ level test cases \end{tabular}& URL (*)\\ \hline
   \hline
\multirow{4}{*}{Bluetooth} & AVRCP: A/V Remote Control Profile 1.6.3 & 170  & 160,783  & {\scriptsize (R-1)} & 118 & {\scriptsize (T-1)}  \\  \cline{2-7}
 & BAP: Basic Audio Profile 1.0.2 & 277  & 305,272  &  {\scriptsize (R-2)}& 119 &  {\scriptsize (T-2)}\\  \cline{2-7}
 & HFP : Hands-Free Profile 1.9 & 143  & 187,508  &  {\scriptsize (R-3)} & 83 &  {\scriptsize (T-3)}\\  \cline{2-7}
 & VDP: Video Distribution Profile 1.1 & 42  & 28,942  &  {\scriptsize (R-4)}  & 38 &  {\scriptsize (T-4)}  \\ \hline
\multirow{4}{*}{Mozilla} & Bookmarks & 11  & 4,078  &  {\scriptsize (R-5)} & 22 &  {\scriptsize (T-5)} \\  \cline{2-7}
 & Theme & 8  & 123  &  {\scriptsize (R-6)} & 19 &  {\scriptsize (T-6)} \\  \cline{2-7}
 & PasswordManager & 5  & 3,991  &  {\scriptsize (R-7)} & 22 &  {\scriptsize (T-7)} \\  \cline{2-7}
 & Browser History & 5  & 800  &  {\scriptsize (R-8)} & 11 &  {\scriptsize (T-8)}  \\ \hline
   \hline
  \end{tabular}

 \begin{tablenotes}
 \scriptsize	
\item URL (*) Feb. 2025 accessed.
\item[(R-1)] \texttt{https://files.bluetooth.com/download/avrcp\_v1-6-3/}
\item[(T-1)] \texttt{https://files.bluetooth.com/download/avrcp-ts-p21-pdf/}
\item[(R-2)] \texttt{https://files.bluetooth.com/download/bap\_v1-0-2/} 
\item[(T-2)] \texttt{https://files.bluetooth.com/download/bap-ts-p6-pdf/} 
\item[(R-3)] \texttt{https://www.bluetooth.org/DocMan/handlers/DownloadDoc.ashx?doc\_id=574215}
\item[(T-3)] \texttt{https://files.bluetooth.com/download/hfp-ts-p27-pdf/} 
\item[(R-4)] \texttt{https://www.bluetooth.org/docman/handlers/DownloadDoc.ashx?doc\_id=260869}
\item[(T-4)] \texttt{https://files.bluetooth.com/download/vdp-ts-p6-pdf/}
\item[(R-5)] \texttt{https://wiki.mozilla.org/Bookmarks\_Use\_Cases}
\item[(T-5)] \texttt{https://www-archive.mozilla.org/quality/browser/front-end/testcases/bookmarks/}
\item[(R-6)] \texttt{https://support.mozilla.org/en-US/kb/use-themes-change-look-of-firefox}
\item[(T-6)] \texttt{https://www-archive.mozilla.org/quality/browser/front-end/testcases/themes/} 
\item[(R-7)] \texttt{https://wiki.mozilla.org/Firefox:Password\_Manager:UI} \\ and \texttt{https://wiki.mozilla.org/Firefox3/Product\_Requirements\_Document},
\item[(T-7)] \texttt{https://www-archive.mozilla.org/quality/browser/front-end/testcases/password-manager/} 
\item[(R-8)] \texttt{https://wiki.mozilla.org/Browser\_History}, 
\item[(T-8)] \texttt{https://www-archive.mozilla.org/quality/browser/front-end/testcases/history/} 

\end{tablenotes}
\end{threeparttable}
\end{table*}


\textbf{GHL-F}: In addition to GHL, test case generation includes combinations of {\bf f}unctions (GHL-F).
\begin{itembox}[l]{GHL-F}
The attached file shows requirements of the system.
Extract as much as possible functions for each section and create combinations of the functions as test cases.
\end{itembox}

\subsection{Datasets}

For the experimental dataset, we utilize publicly available documents and their corresponding test cases. The datasets that meet these criteria include those from 
Bluetooth \cite{bluetooth} and Mozilla \cite{mozilla-requirements, mozilla-testcases} as shown in Table \ref{table:datasets}. 
Table \ref{table:datasets} presents the number of requirements, pages, words, high-level test cases, and source information for each dataset.

Bluetooth is one of the short-range wireless communication standards for digital devices. It is used for exchanging information between devices located within a range of several to tens of meters using radio waves. This technology is applied in various everyday devices, such as audio-video remote controls and hands-free headsets. For these devices, the respective requirement specifications and corresponding test cases are publicly available.
From these publicly available requirement documents and test cases, we use four types in this experiment: AVRCP: A/V Remote Control Profile, BAP: Basic Audio Profile, HFP: Hands-Free Profile, and VDP: Video Distribution Profile.

The notation for test cases is defined in the Bluetooth testing strategy as follows: 
$<$spec abbreviation$>$/$<$IUT role$>$/$<$class$>$/$<$feat$>$/$<$func$>$/$<$subfunc$>$/$<$cap$>$/$<$xx$>$-$<$nn$>$-$<$y$>$.
For instance, it is written as AVRCP/CT/CON/BV-01-C. This indicates a test for the "Valid Behavior (BV)" of the "Connection Establishment and Release for Browsing" (CON) function in the Controller Role (CT) within the A/V Remote Control Profile.

The term "Mozilla" refers to internet-related software developed by the former Netscape Communications Corporation. Notable examples include Firefox and Thunderbird. The requirement documents and test cases for these software applications have been made publicly available. In the present study, we utilized the requirement documents related to Firefox, specifically focusing on four features: Bookmarks, Themes, Password Manager, and Browser History.

\subsection{Evaluation Methodology}

Initially, the experiment is conducted by generating test cases using a LLM for each requirement document in the datasets. This process is repeated three times for the baseline, GHL, and GHL-F. As a result, test cases are generated for four requirement documents in the Bluetooth dataset and four requirement documents in the Mozilla dataset, each repeated three times. To confirm practical applicability, the time required for each process is also measured. Each dataset contains truth test cases described in natural language. The evaluation is performed by comparing these truth test cases with the generated test cases. 
For the comparative evaluation method, the descriptions of the truth test cases, and the generated test cases are compared by semantic similarity.  We employ the semantic similarity as vectorizing \cite{1301.3781} by the LLM and calculating cosine similarity \cite{10.5555/866292, reimers-gurevych-2019-sentence} between the vectors.
Generally, a similarity score of 0.7 or higher is considered identical \cite{huang2008similarity, manning2009introduction,turney2010frequency}, 
and in this study, a threshold of 0.7 was used to determine equivalence.

\renewcommand{\arraystretch}{1.2}

\begin{table*}[h!]
 \caption{Examples of High similarity between Truth and Generated test cases}
 \label{table:high-sim}
 \centering
  \begin{tabular}{l|p{1.1cm}|p{6.2cm}|p{6.2cm}|r}
   \hline
Dataset & Function & Truth: High-level test cases  & Generated by GHL-F: High-level test cases  & Similarity \\ \hline
   \hline
Mozilla & Password Manager & Verify that when you have saved a username and password for a website and you later change the password, that password manager picks up the new password. & Verify that the password manager correctly updates saved passwords when the user changes their password on the site. & 0.81  \\ \cline{3-5}
 &  & Verify that the master password is visually distinctive from other password prompts (indicates the functionality and importance of the master password). & Verify that the color of the Master Password popup is distinct from other popups to avoid confusion. & 0.79  \\ \cline{2-5}
 & Bookmarks & Verify that bookmarks are properly deleted from bookmarks file and disappear from all access points. & User deletes a bookmark and checks that it no longer appears in the bookmarks list. & 0.72  \\ \cline{3-5}
 &  & Verify that several bookmarks are deleted at one time. & User deletes a bookmark and checks that it is removed from all locations (bookmarks menu). & 0.69  \\ \cline{2-5}
 & History & Verify that the correct link is recorded after a server redirect. & Verify that the 'moz\_history' table can store a new entry with a unique ID & 0.71  \\ \cline{3-5}
 &  & Verify that browser remembers scroll state when returning to pages in SH. & Verify that the 'moz\_historyvisit' table can store a new entry with a source visit ID & 0.60  \\ \cline{2-5}
 & Themes & Test all the items under Apply Theme under View menu. Selecting Theme Preferences. Get New Themes. Classic. Modern. & Switch themes and verify the appearance changes & 0.62  \\ \cline{3-5}
 &  & Verify if themes exist under View menu & Install a theme and verify it is enabled & 0.61  \\ \hline
Bluetooth & AVRCP & AVRCP/CT/CON/BV-01-C & AVRCP/CT/CON/BV-01-C & 1.00  \\ \cline{3-5}
 &  & AVRCP/TG/MPS/BI-01-C & AVRCP/TG/MPS/BI-01-C & 1.00  \\ \cline{2-5}
 & BAP & BAP/UCL/DISC/BV-05-C & BAP/UCL/DISC/BV-05-C & 1.00  \\ \cline{3-5}
 &  & BAP/USR/DISC/BV-07-C & BAP/USR/DISC/BV-07-C & 1.00  \\ \cline{2-5}
 & HFP & HFP/HF/ACS/BV-01-C & HFP/HF/ACS/BV-01-C & 1.00  \\ \cline{3-5}
 &  & HFP/AG/ACS/BV-02-C & HFP/AG/ACS/BV-02-C & 1.00  \\ \cline{2-5}
 & VDP & VDP/SNK/SYN/BV-01-C & VDP/SNK/SYN/BV-01-C & 1.00  \\ \cline{3-5}
 &  & VDP/SRC/HC/BV-02-C & VDP/SRC/HC/BV-02-C & 1.00  \\ \cline{3-5}
   \hline
  \end{tabular}
\end{table*}

\begin{table*}[h!]
 \caption{An Example of the similarity between the truth test case and the generated test cases: \\In the case of Mozilla's Password Manager function}
 \label{table:similarity}
 \centering
  \begin{tabular}{p{27em}|p{27em}|r}
   \hline
Truth: High-level test case & Generated by GHL-F: High-level test cases & similarity \\ \hline
\hline
& Verify that the password manager correctly updates saved passwords when the user changes their password on the site. & 0.808 \\ \cline{2-3}
\multirow{7}{*}{\begin{tabular}{l} Verify that when you have saved a username and password \\ for a website and you later change the password, that password \\ manager picks up the new password \end{tabular}} & Verify that the password manager can provide feedback to the user when a password is saved or updated successfully. & 0.701 \\\cline{2-3}
& Verify that the password manager handles multiple accounts for the same site correctly in a live environment. & 0.651 \\ \cline{2-3}
& Verify that the password manager does not allow saving passwords without entering the Master Password when it is enabled. & 0.644 \\ \cline{2-3}
& Verify that the notification bar for saving passwords is visually distinct and easily noticeable. & 0.500 \\ \cline{2-3}
& Verify that the notification bar appears correctly after a successful login. & 0.314 \\ \cline{2-3}
& Verify that the search functionality does not crash or hang when searching with a very long keyword. & 0.166 \\ \cline{2-3}
   \hline
  \end{tabular}
\end{table*}

\begin{table*}[h]
 \caption{Summary of the experiments zero-shot, Proposed method 1 and 2 for the datasets}
 \label{table:summary}
 \centering
  \begin{tabular}{l|ll|>{\raggedleft\arraybackslash}p{1.8cm}|>{\raggedleft\arraybackslash}p{1.8cm}|>{\raggedleft\arraybackslash}p{1.8cm}}
   \hline
 &  &  & zero-shot & GHL & GHL-F \\ 
\hline
\hline
Mozilla & (A) & num\_truth & 19  & 19  & 19  \\ \cline{2-6}
 & (B) & num\_gen\_ts & 16  & 108  & 117  \\ \cline{2-6}
 & (C) & num\_mat\_uniq\_truth & 0  & 4  & 7  \\ \cline{2-6}
 & (D) & num\_mat\_uniq\_gen & 0  & 8  & 19  \\ \cline{2-6}
 & (B)$\div$(A) & ratio\_num & 0.89  & 5.82  & 6.32  \\ \cline{2-6}
 & (D)$\div$(B) Macro Precision & ratio\_mat\_gen & 0.02  & 0.07  & 0.17  \\ \cline{2-6}
 & (C)$\div$(A) Macro Recall & ratio\_mat\_truth & \underline{{\bf 0.02}}  & \underline{{\bf 0.20}}  & \underline{{\bf 0.37}}  \\ \cline{2-6}
 & F1 -score &  & 0.02  & 0.10  & 0.23  \\ \cline{2-6}
 &  & Duration time (hh:mm:ss) & 0:01:51 & 0:13:46 & 0:12:34 \\ \hline
\hline
Bluetooth & (A) & num\_truth & 90  & 90  & 90  \\ \cline{2-6}
 & (B) & num\_gen\_ts & 29  & 51  & 98  \\ \cline{2-6}
 & (C) & num\_mat\_uniq\_truth & 54  & 57  & 73  \\ \cline{2-6}
 & (D) & num\_mat\_uniq\_gen & 25  & 38  & 56  \\ \cline{2-6}
 & (B)$\div$(A) & ratio\_num & 0.35  & 0.82  & 1.42  \\ \cline{2-6}
 & (D)$\div$(B) Macro Precision & ratio\_mat\_gen & 0.83  & 0.75  & 0.60  \\ \cline{2-6}
 & (C)$\div$(A) Macro Recall & ratio\_mat\_truth & \underline{{\bf 0.65}}  & \underline{{\bf 0.69}}  & \underline{{\bf 0.84}}  \\ \cline{2-6}
 & F1 -score &  & 0.73  & 0.72  & 0.70  \\ \cline{2-6}
 &  & Duration time (hh:mm:ss) & 0:02:23 & 0:20:10 & 0:19:43 \\ \hline
   \hline
  \end{tabular}
\end{table*}

\subsection{Experimental Environment}

In the experimental environment, OpenAI's GPT-4o Small \cite{2409.11547} was used. This model was chosen due to its relatively high token limit. Python 3.9 was used as the programming language. When using OpenAI's API, the seed was fixed, and the temperature was set to 0. Furthermore, for creating the embedding vectors used to compare the generated test case descriptions with the truth test case descriptions, OpenAI's text-embedding-3-small \cite{text-embedding-3-small} model was employed.

\section{Experimental Results and Evaluation}

Table \ref{table:high-sim} shows the generated test cases that exhibit a high degree of similarity to the truth test cases, based on the dataset and each specific feature. Upon comparing the truth test cases with the generated ones, it becomes evident that the terminology and the aspects to be verified within the test cases are similar. In the case of Bluetooth test cases, they are described following the test case notation, where features, classes, sub-features, as well as test cases for both successful and failure scenarios, are delineated and separated by slashes.

Table \ref{table:similarity} presents a comparison between the truth test cases based on the Mozilla Password Manager requirements document and the generated test cases. The similarity was calculated by vectorizing each sentence using embedding vectors and computing the cosine similarity. The similarity values range from 0.81 to 0.17, and the cases are approximately selected and arranged in increments of 0.1 from most like least similar.

In examining the case with the highest similarity of 0.81, it is observed that terms such as "password," "correctly," and "change" semantically match between the generated and the truth test case descriptions, indicating a highly similar description. The similarity threshold is set at 0.7, rounded to 0.7 at the second decimal place. Therefore, a similarity of 0.65 or higher is considered a match. Near this threshold, in the comparison with a similarity of 0.65, terms like "password" and "correctly" match, but terms such as "multiple accounts" differ. Conversely, in the least similar comparison with a similarity of 0.17, no matching terms are observed between the generated and truth test case descriptions.

Additionally, it is noteworthy that the "Generated" test cases cover a wide range of aspects, including user interface operability, notification visibility, impact on performance, and error handling of the user interface. Particularly, cases with low similarity (0.31 or below) are likely to be test cases unrelated to password updates.

Table \ref{table:summary} compares the performance of three testing techniques: zero-shot, GHL and GHL-F—across two datasets: Mozilla and Bluetooth. For each testing technique, the following metrics were measured: num\_truth (A), the number of correct data; num\_gen\_ts (B), the number of generated test cases; num\_mat\_uniq\_truth (C), the number of uniquely matched correct data test cases; and num\_mat\_uniq\_gen (D), the number of uniquely generated test cases. Additionally, the following metrics were calculated from these: ratio\_num (D)$\div$(A), the generation rate (the proportion of generated cases relative to the correct data); ratio\_mat\_gen (D)$\div$(B), precision (the proportion of generated cases corresponding to the correct data); ratio\_mat\_truth (C)$\div$(A), recall (the proportion of matches within the correct data); and the F1-score, the harmonic mean of precision and recall.

For the Mozilla dataset, the results indicate that the zero-shot method yields a very low recall (ratio\_mat\_truth) of 0.02, a high generation rate (ratio\_num) of 0.89, but a very low precision (ratio\_mat\_gen) of 0.02. The test generation time is short at 0:01:51. GHL improves recall to 0.20 and achieves a very high generation rate of 5.82, though precision remains low at 0.07. The test generation time increases to 0:13:46. This method requires executing prompts for each test design technique, resulting in time consumption proportional to the number of techniques. GHL-F further improves recall to 0.37, maintains a high generation rate of 6.32, and improves precision to 0.17. The test generation time is the longest at 0:19:30.

For the Bluetooth dataset, the zero-shot method achieves a high recall of 0.65, a moderate generation rate of 0.35, and a short test generation time of 0:02:23. GHL slightly improves recall to 0.69, increases the generation rate to 1.42, and achieves a precision of 0.55. The test generation time is longer at 0:20:10. GHL-F achieves the highest recall of 0.80, increases the generation rate to 1.56, and improves precision to 0.62. The test generation time is slightly reduced to 0:19:43.

For the Mozilla dataset, Proposed method 2 achieves the highest recall (0.37) but also increases the generation rate and test generation time. The zero-shot method is fast but has a very low recall. Conversely, for the Bluetooth dataset, GHL-F demonstrates the highest recall (0.80) and precision (0.62), indicating superior performance. While the zero-shot method achieves high recall, it generates fewer cases and is efficient, but its low precision remains a challenge.

\begin{table*}[h]
 \caption{Detail results of GHL-F for the dataset Mozilla}
 \label{table:mozilla_results}
 \centering
  \begin{tabular}{ll|>{\raggedleft\arraybackslash}p{1.3cm}|>{\raggedleft\arraybackslash}p{1.3cm}|>{\raggedleft\arraybackslash}p{1.3cm}|>{\raggedleft\arraybackslash}p{1.3cm}||>{\raggedleft\arraybackslash}p{1.3cm}}
   \hline
 &  & Bookmarks & History & Password Manager & Themes & Average \\ \hline
  \hline
 (A) &  num\_truth & 22  & 19  & 22  & 11  & 19  \\ \hline
 (B) &  num\_gen\_ts & 96  & 130  & 192  & 49  & 117  \\ \hline
 (C) &  num\_mat\_uniq\_truth & 4  & 4  & 19  & 0  & 7  \\ \hline
 (D) &  num\_mat\_uniq\_gen & 11  & 6  & 61  & 0  & 19  \\ \hline
 (B)$\div$(A) &  ratio\_num & 4.36  & 6.86  & 8.71  & 4.48  & 6.32  \\ \hline
 (D)$\div$(B) Macro Precision &  ratio\_mat\_gen & 0.10  & 0.05  & 0.34  & 0.00  & 0.17  \\ \hline
 (C)$\div$(A) Macro Recall &  ratio\_mat\_truth & 0.17  & 0.23  & 0.88  & 0.00  & \underline{{\bf 0.37}}  \\ \hline
   \hline
  \end{tabular}
\end{table*}

\begin{table*}[h]
 \caption{Detail results of GHL-F for the dataset Bluetooth}
 \label{table:bluetooth_results}
 \centering
  \begin{tabular}{ll|>{\raggedleft\arraybackslash}p{1.3cm}|>{\raggedleft\arraybackslash}p{1.3cm}|>{\raggedleft\arraybackslash}p{1.3cm}|>{\raggedleft\arraybackslash}p{1.3cm}||>{\raggedleft\arraybackslash}p{1.3cm}}
   \hline
  &  & AVRCP & BAP & HFP & VDP & Average \\ \hline
   \hline
 (A) &  num\_truth & 118  & 119  & 83  & 38  & 90  \\ \hline
 (B) &  num\_gen\_ts & 86  & 69  & 126  & 109  & 98  \\ \hline
 (C) &  num\_mat\_uniq\_truth & 101  & 82  & 73  & 35  & 73  \\ \hline
 (D) &  num\_mat\_uniq\_gen & 56  & 23  & 71  & 76  & 56  \\ \hline
 (B)$\div$(A) &  ratio\_num & 0.73  & 0.58  & 1.52  & 2.86  & 1.42  \\ \hline
 (D)$\div$(B) Macro Precision &  ratio\_mat\_gen & 0.65  & 0.33  & 0.66  & 0.74  & 0.60  \\ \hline
 (C)$\div$(A) Macro Recall &  ratio\_mat\_truth & 0.86  & 0.69  & 0.88  & 0.93  & \underline{{\bf 0.84}}  \\ \hline
   \hline
  \end{tabular}
\end{table*}

Tables \ref{table:mozilla_results} and \ref{table:bluetooth_results} present the average values based on various metrics, including "num\_truth," "num\_gen\_ts," "num\_mat\_uniq\_truth," and "num\_mat\_uniq\_gen." These metrics are crucial for evaluating the quality and diversity of the generated data.
For instance, the average data reveals the uniqueness and truthfulness ratio of the generated data. Specifically, the calculated ratios provide indicators of the extent to which the generated data is based on truth. This offers a benchmark for evaluating the performance of the generative model.
Additionally, metrics such as macro precision and macro recall are also presented, which are important for assessing the overall performance of the generative model. In particular, macro precision indicates the accuracy of the generated data, while macro recall reflects how well the generated data captures the truth.
As demonstrated, the data in the attached files provide significant information for evaluating the performance of generative models. Utilizing these metrics is expected to contribute to the improvement of generative models in future research and practice.

\begin{table}[h]
 \caption{Extracted test design techniques from the dataset Bluetooth}
\vspace{-0.5\baselineskip}
 \label{table:bluetooth_tecs}
 \centering
  \begin{tabular}{l|p{2em}p{2em}p{2em}p{2em}|p{2em}}
   \hline
Test design techniques &  VDP & HFP & AVRCP & BAP & Sub-total \\ \hline
   \hline
Boundary Value Analysis & 1 & 1 & 1 & 1 & 4 \\ \hline
Decision Table Testing & 1 & 1 & 1 & 1 & 4 \\ \hline
Equivalence Partitioning & 1 & 1 & 1 & 1 & 4 \\ \hline
Error Guessing & 1 & 1 & 1 & 1 & 4 \\ \hline
Exploratory Testing & 1 & 1 & 1 & 1 & 4 \\ \hline
State Transition Testing & 1 & 1 & 1 & 1 & 4 \\ \hline
Use Case Testing & 1 & 1 & 1 & 1 & 4 \\ \hline
Performance Testing & 1 & 1 & 1 &  & 3 \\ \hline
Regression Testing &  & 1 & 1 &  & 2 \\ \hline
Security Testing &  & 1 & 1 &  & 2 \\ \hline
Ad-hoc Testing &  &  &  & 1 & 1 \\ \hline
Compatibility Testing &  & 1 &  &  & 1 \\ \hline
Integration Testing &  & 1 &  &  & 1 \\ \hline
Model-Based Testing &  &  &  & 1 & 1 \\ \hline
Risk-Based Testing & 1 &  &  &  & 1 \\ \hline
System Testing &  & 1 &  &  & 1 \\ \hline
User Acceptance Testing &  & 1 &  &  & 1 \\ \hline
   \hline
Total & 9 & 14 & 10 & 9 & 42 \\ \hline
  \end{tabular}
\vspace{5mm}
\\
 \caption{Extracted test design techniques from the dataset Mozilla}
\vspace{-0.5\baselineskip}
 \label{table:mozilla_tecs}
 \centering
  \begin{tabular}{l|p{2em}p{2em}p{2em}p{2em}|p{2em}}
   \hline
Test design techniques & Book-marks & History & Pass-ward Manager & Them-es & Sub-total \\ \hline
   \hline
Boundary Value Analysis & 1 &  & 1 & 1 & 3 \\ \hline
Equivalence Partitioning & 1 &  & 1 & 1 & 3 \\ \hline
Performance Testing & 1 & 1 &  & 1 & 3 \\ \hline
Exploratory Testing & 1 &  & 1 &  & 2 \\ \hline
State Transition Testing & 1 &  &  & 1 & 2 \\ \hline
Use Case Testing & 1 &  & 1 &  & 2 \\ \hline
Compatibility Testing &  &  & 1 &  & 1 \\ \hline
Data Integrity Testing &  & 1 &  &  & 1 \\ \hline
Database Testing &  & 1 &  &  & 1 \\ \hline
Decision Table Testing & 1 &  &  &  & 1 \\ \hline
Negative Testing &  & 1 &  &  & 1 \\ \hline
Query Testing &  & 1 &  &  & 1 \\ \hline
Random Testing &  &  & 1 &  & 1 \\ \hline
Regression Testing &  &  & 1 &  & 1 \\ \hline
Requirement-based Testing &  &  &  & 1 & 1 \\ \hline
Scenario Testing &  &  & 1 &  & 1 \\ \hline
Session Testing &  & 1 &  &  & 1 \\ \hline
Usability Testing &  & 1 &  &  & 1 \\ \hline
User Acceptance Testing &  &  & 1 &  & 1 \\ \hline
   \hline
Total & 7 & 7 & 9 & 5 & 28 \\ \hline
   \hline
  \end{tabular}
\end{table}

Table \ref{table:mozilla_results} shows the results of applying the proposed method to each requirement document (Bookmarks, History, Password Manager, Theme) in the Mozilla dataset to create test cases. The values averaged over these four requirement documents are presented as the results for Mozilla. 
For instance, regarding the bookmark feature, 96 test cases were generated against 22 truth test cases. Among the 22 truth test cases, 4 matched the generated test cases. This represents that out of the 96 generated test cases, 11 were correct. Note that, the comparison of test cases considers a similarity of 0.7 or higher as a match, and a single truth test case may match multiple generated test cases, which accounts for the discrepancy between the 4 matches with truth test cases and the 11 correct generated test cases. Given that 96 test cases were generated for 22 correct ones, this results in 4.36 times the number of test cases being generated. Since 11 out of the 96 generated test cases were correct, the macro precision is 0.10. Since 4 out of the 22 truth test cases were matched, the macro recall is 0.17.

Table \ref{table:bluetooth_results} presents the results of applying the proposed method to create test cases for each requirement document, namely AVRCP, BAP, HFP, and VDP, in the Bluetooth dataset. The average of the results for these four requirement documents is presented as the result for Bluetooth. For instance, for AVRCP, the total number of truth test cases is 118. The proposed method generated 86 test cases, of which 101 corresponded to truth test cases. 
Note that the comparison of test cases considers a match when the similarity is 0.7 or higher. Consequently, a single generated test case may match multiple correct test cases. In this study, 86 generated test cases matched with 101 correct test cases.
Of the 86 generated test cases, 56 matched the correct values. Based on these numbers, 86 test cases were generated, which is 0.73 times the 118 correct ones. The precision and recall are calculated to be 0.65 and 0.86, respectively.

In this paper, we anticipate the future automation of test case generation, which allows us to offset the increase in the number of test cases through automated execution, even if the number of test cases increases. Therefore, we do not place significant importance on the decline in precision. On the other hand, we emphasize the importance of recall, as we focus on how many truth test cases can be identified.

For instance, in experiments using the Bluetooth dataset, the number of generated test cases was 29 for zero-shot, 51 for GHL, and 98 for GHL-F. Corresponding precision values decreased to 0.83, 0.75, and 0.60, respectively, but these values are not considered critical. Instead, we emphasize the improvement in recall values, which increased from 0.65 to 0.69, and further to 0.84.

Table \ref{table:bluetooth_tecs} shows the results of extracting test design techniques from requirement documents using a LLM based on the proposed method. Specifically, when extracting test design techniques from the VDP requirement document using an LLM, techniques ranging from boundary value analysis and decision table testing to performance testing and risk-based testing were extracted. In the case of VDP, a total of 9 types of test design techniques were extracted. Similarly, Table \ref{table:mozilla_tecs} shows the results of extracting test design techniques in the Mozilla dataset. For the Bookmarks requirement document, techniques such as boundary value analysis, equivalence partitioning, use case testing, and decision table testing were extracted. It is shown that a total of 7 types of test design techniques were extracted for the Bookmarks requirement. Table \ref{table:bluetooth_tecs} shows the results of extracting test design techniques in the Bluetooth dataset. The extracted test design techniques are listed in order of frequency. In both the Bluetooth and Mozilla datasets, boundary value analysis was most frequently extracted. Additionally, it is evident that test design techniques described in the international standard ISO/IEC/IEEE 29119 Part 4, such as system testing and user acceptance testing, were extracted.

\begin{figure}[t]
\centering
 \includegraphics[width=90mm,bb=0 0 606 375]{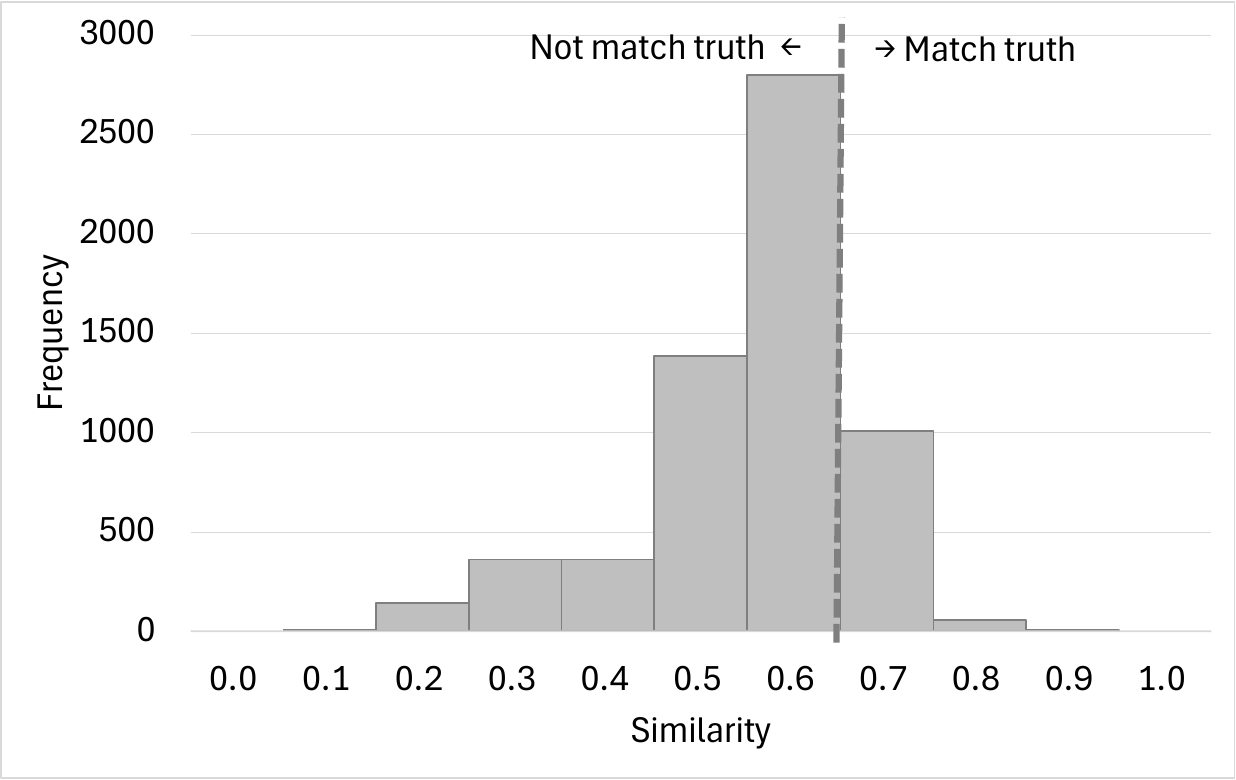}
 \vspace*{-0.5cm}
 \caption{Distribution of similarity for generated test cases when recall is high (Password Manager feature)}
\label{figure:hist_password}
\vspace{5mm}

\centering
 \includegraphics[width=90mm,bb=0 0 606 375]{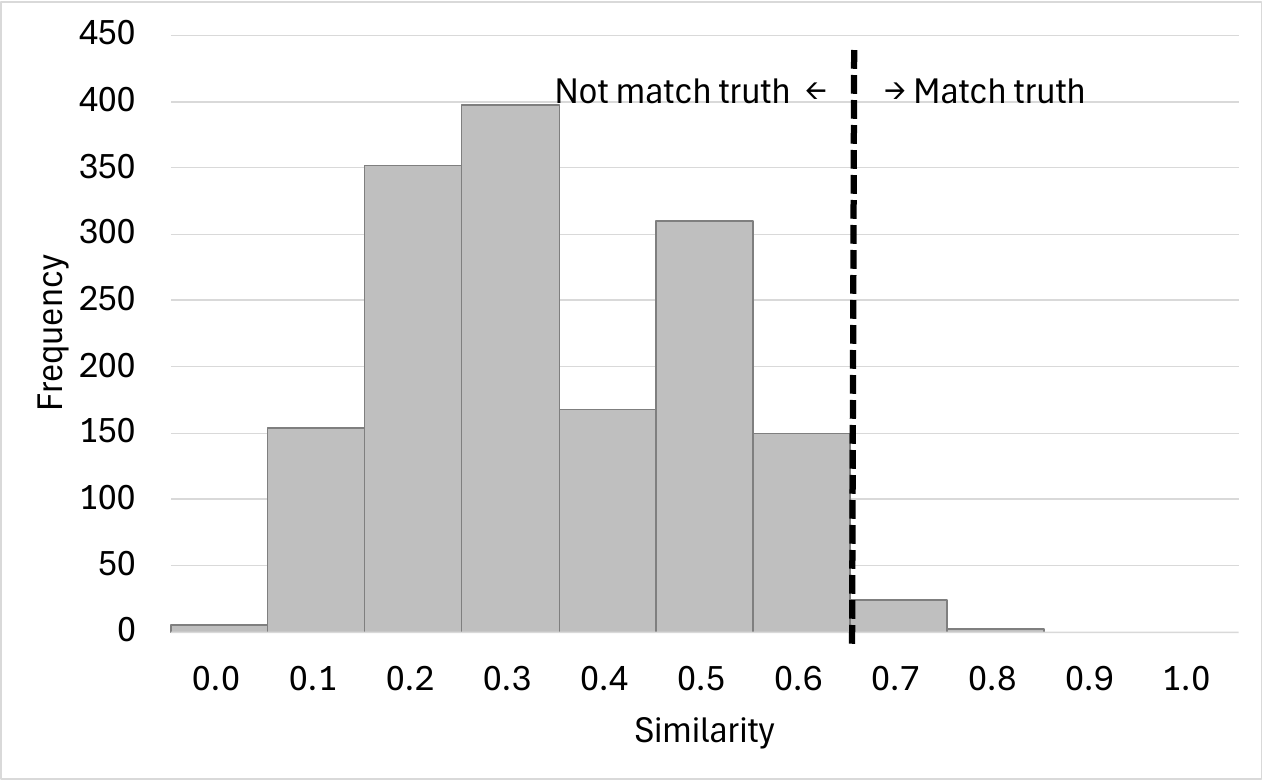}
 \vspace*{-0.5cm}
 \caption{Distribution of similarity for generated test cases when recall is low (Bookmarks feature)}
\label{figure:hist_bookmarks}
\end{figure}

Fig. \ref{figure:hist_password} and Fig. \ref{figure:hist_bookmarks} illustrate the distribution of similarity between the descriptions of truth test cases and the generated test cases, as presented in Table \ref{table:similarity}. Fig. \ref{figure:hist_password} shows the similarity distribution of test cases generated from the requirement documents of the password manager in the Mozilla dataset. On average, 192 test cases were generated, with 22 being the truth test cases. For each of the 192 generated test cases, the similarity to each truth test case was calculated, resulting in an average total of approximately 4,224 similarity scores. 
In the figure, the region to the right of the dashed line indicating the similarity threshold of 0.7 is considered to match the truth test cases. Conversely, the region to the left of the dashed line at the similarity threshold of 0.7 indicates that there is no match with any truth test cases. Analyzing the proportion of correct and incorrect cases in this figure reveals that the proportion of incorrect cases is significant. This study emphasizes the importance of finding correct cases, as it assumes the automation of test execution using generated test cases. Although incorrect or redundant test cases exist, it is assumed that these can be efficiently handled through automation.
The distribution of these scores is shown, indicating that the password manager yielded high similarity results. Observing this distribution, a peak at a relatively high position can be confirmed. Conversely, Fig. \ref{figure:hist_bookmarks} shows the similarity distribution of test cases generated from the requirement documents of the bookmarks in the Mozilla dataset. Observing this distribution reveals two significant peaks in the similarity histogram. As a result, the overall similarity for bookmarks is shown to be lower.

\vspace{-0.5\baselineskip}
\section{Discussion}
In accordance with the points elucidated in this paper, we will conduct a discussion of the experimental results.

{\bf RQ1: What type of prompt is effective for generating high-level test cases from requirement documents using LLMs with prompts alone?}

The experimental results demonstrated that our proposed method enables the generation of high-level test cases from requirement documents. The macro averages were 0.84 and 0.37, respectively, indicating that high-level test cases could be obtained with a certain degree of accuracy. The prompt used in this study first extracts test design techniques and then generates test cases from the requirement documents based on each test design technique. This prompt yielded the most favorable results. It is considered that high-level test cases can be created using this prompt. As shown in previous research, this prompt is a step-by-step prompt. In this regard, our results are consistent with the findings of prior studies. Our novelty lies in adding the element of extracting test design techniques to the step-by-step prompt. This allowed us to address Research Question 1 (RQ1) of this study.

{\bf RQ2: How can the validity of the generated high-level test case descriptions be evaluated?}

The validity evaluation of the generated high-level test cases was conducted by vectorizing them using embedding vectors and calculating the semantic similarity with the truth test cases. The threshold for the validity determination was set at 0.7, based on previous research. 
The experimental results shown in Table \ref{table:similarity} confirmed through expert evaluation that there were no issues with the similarity of the high-level test case descriptions. 
From these results, it is considered that the validity evaluation regarding the similarity of test cases is ensured. As a method for evaluating the effectiveness of test cases, mutation testing \cite{5487526} exists. This method evaluates the effectiveness of test cases by intentionally embedding bugs into an existing system and checking whether the test cases can detect those bugs. Specifically, if a bug is detected, the test case is deemed effective; if not, it is deemed ineffective. Although this method was not used in this study, we intend to consider it as a method for determining the best cases in future research.

{\bf RQ3: What are the current achievements and future challenges for the practical implementation of automatic high-level test case generation from requirements?}

The goal of automatically generating high-level test cases from requirements is to improve efficiency and prevent omissions in test cases. In this regard, the macro-recall of 0.84 in this experiment suggests that automatic generation of high-level test cases is feasible. 
The time required for generating high-level test cases from requirements using a LLM is approximately 20 minutes for around 100 cases. This is significantly less than the 500 minutes it would take a human to create the same number of test cases, assuming an average of 5 minutes per case. This efficiency highlights the potential of the proposed method to motivate further automation in test case generation in the future.
This study focused on the generation of high-level test cases. If automation is to be considered in future processes, it will be necessary to automatically execute tests using the generated test cases. At that time, it will be necessary to clarify the expected results during test execution. This remains a challenge for future work.

\vspace{-0.5\baselineskip}
\section{Conclusion}
In this paper, we propose a method for generating high-level test cases from requirement documents using only prompts, without employing RAG. Currently, RAG has been frequently used for generating high-level test cases; however, applying RAG requires structuring domain-specific knowledge, which is labor-intensive. In our research, we explored a method to generate test cases without RAG by inputting requirement documents into a LLM and generating test design techniques corresponding to the requirements.

We validated the effectiveness of the proposed method through experiments using datasets from Bluetooth and Mozilla. The results showed that we achieved recall rates of 0.81 for the Bluetooth dataset and 0.37 for the Mozilla dataset, demonstrating that high-level test cases can be generated with a certain degree of accuracy. This indicates that a method not utilizing RAG can also be practical.
For the evaluation of the generated test cases, we employed similarity calculations using embedding vectors. We set a criterion that considers a similarity of 0.7 or higher as identical and confirmed through expert evaluation that there were no issues. Future challenges include the automatic execution of tests using the generated test cases and the clarification of expected outcomes.

The results of this study contribute to the efficiency of test case generation in the software industry and represent a step towards establishing a universally applicable automatic generation method for diverse requirement documents. In the future, we plan to conduct further validation using more diverse datasets and advance research on the automation of test execution.


\bibliographystyle{IEEEtran}
\bibliography{llm_hl_testgen} 

\end{document}